# A Perfect Specialization Model for Gravity Equation in Bilateral Trade based on Production Structure

**Majid Einian**[*]
Graduate School of Management and Economics, Sharif University of Technology, Iran
**Farshad Ranjbar Ravasan**
Paris School of Economics, Paris, France

**Abstract**
Although initially originated as a totally empirical relationship to explain the volume of trade between two partners, gravity equation has been the focus of several theoretic models that try to explain it. Specialization models are of great importance in providing a solid theoretic ground for gravity equation in bilateral trade. Some research papers try to improve specialization models by adding imperfect specialization to model, but we believe it is unnecessary complication. We provide a perfect specialization model based on the phenomenon we call tradability, which overcomes the problems with simpler initial. We provide empirical evidence using estimates on panel data of bilateral trade of 40 countries over 10 years that support the theoretical model. The empirical results have implied that tradability is the only reason for deviations of data from basic perfect specialization models.

**Keywords:** Bilateral Trade, Gravity Equation, Perfect Specialization, Tradability.

**JEL Classification:** F11, F14, C23, E23.

---

[*] Corresponding Aauthor, Email: einian85@gmail.com.

# 1. Introduction

Although trade is as old as economics itself since Ricardo (1817), it is gaining much more importance as international trade has been growing tremendously. Gravity equation is a form of empirical relation explaining the flow of bilateral trade by the size of two engaging countries and negatively by distance between them usually in a form resembling the law of gravity in Physics. The traditional relationship did not have a theoretical basis, but theories of trade have tried to explain this equation (Deardorff, 1998).

There are a lot of theories which explain the patterns of international trade. A class of these theories is based on relative factor abundance. One of the common relative-factor-abundance models is the Heckscher-Ohlin model. This theory predicts that trade patterns would be based on relative factor advantages. Those countries with a relative abundance of one factor are expected to produce goods that require a relatively large amount of that factor in their production. Although this model generally is accepted as the theory of trade but does not satisfy empirical results (Bergstrand, 1989).

A study by Wassily Leontief indicates that the exports of United State as the most capital endowed country include more labor intensive commodities, which suggests the opposite result. This contradiction is known as the Leontief paradox. The Leontief paradox makes doubt about that Heckscher-Ohlin works in real world.

An alternative theory, the first which was proposed by Linder (1961), claims that the pattern of trade is determined by similarity of two country's preferences (Bohman and Nilson, 2007). Countries with similar demand develop similar industries that result in producing similar products. These countries continue trade in differentiated but similar goods. Linder (1961) writes "The more similar the demand structure of the two countries the more intensive potentially is the trade between these two countries." Importance of Linder's hypothesis considering demand part is what departs this theory from neoclassical theories of trade which pay attention only to production features part. Linder suggests that per capita income can be used as a proxy for preferences. The hypothesis can then be tested by comparing per capita income between trading partners. It means the more similar two country's GDP's are, the more they trade. That result is consistent with the gravity equation.

Helpman and Krugman (1985) develop the Lender's idea. They observed based on the gravity model countries with similar levels of income have been shown to trade more (Bohman and Nilson, 2006). This is not supported by Heckscher-Ohlin model of trade and comparative advantage theory. They introduced Increasing Returns to Scale as fundamental factor that account for part of trade known as intra industry trade (IIT). They relax the neoclassical assumption, perfect competition market. Substantial theoretical progress has been made using three different approaches. These are the Marshallian approach, where economies of scale are assumed external to firms; the Chamberlinian approach, where imperfect competition takes the relatively tractable form of monopolistic competition; and the Cournot approach of non-cooperative quantity-setting firms.

The reciprocal dumping model – in which both countries export the same good to each other to gain higher profits by supplying their product to the other country with lower prices than their own market (Krugman and Obstfeld 2009) – also explains gravity equation. Feenstra et al. (1998) provided evidence for reciprocal dumping by assessing the "home market effect" in separate gravity equations for differentiated and homogeneous goods. The home market effect showed a relationship in the gravity estimation for differentiated goods, but showed the inverse relationship for homogeneous goods. The authors show that this result matches the theoretical predictions of reciprocal dumping playing a role in homogeneous markets.

At all, the literature of gravity model of trade includes two debates: first what model is the theoretical base of gravity equation and second what factors account for deviation of real bilateral trade from gravity form. To answer the first question, Deardorff (1998) claimed that the basic gravity model can be derived from Heckscher-Ohlin as well as the Linder and Helpman-Krugman hypotheses. Deardorff (1998) concludes that, considering how many models can be tied to the gravity model equation, it is not useful for evaluating the empirical validity of theories. Barriers, Demand structure, imperfect specialization are three factors which were noticed as basic factors for deviation from gravity equation.

To answer the second question Evenett and Keller (2002) suggest that relaxing the perfect specialization assumption produces much better results. They support an imperfect specialization based on a model identification approach consisting of two conditions, first the model should provide a regression coefficient less than one, i.e. can match the real-world data; and second the model should be consistent with the correlation of specialization index and the regression coefficient.

As we mentioned the first condition is a kind of gravity equation support identification and the second one holds if the model can provide an explanation for real bilateral trade deviations from traditional gravity equation.

We provide a perfect specialization model based on the phenomenon we call tradability, which explains the less-than-one coefficient by non-tradable share of GDP rather than levels of specialization. We provide empirical evidence using estimations on panel data of bilateral trade of 40 countries over 10 years that support the theoretical model. We also provide some empirical evidence on how imperfect specialization might not address the fundamental deviation factor since high correlation of specialization index with other important deviation factors like trade cost and barriers.

Remainder of the paper is structured as follows. In Section 2 we review Evenett and Keller's model identification approach. Then we introduce our model of perfect specialization based on tradability phenomenon in Section 3. Section 4 provides information about data used in the study and also the tradability index we calculated for 40 countries. Section 5 gives the empirical test results, and Section 6 concludes.

## 2. Model Identification Approach

Evenett and Keller's identification approach consists of two steps based on a regression of this type:

$$X_{ab} = \alpha \left( X_{ab}^{Model} \right) + \varepsilon_{ab} \quad (5)$$

where $X_{ab}^{Model}$ is the trade predicted based on gravity equation.

If we ignore the theoretic base of each specialization model, Evenett and Keller suggest that all perfect specialization models will lead in a gravity equation as in follows:

$$X_{ab} = \frac{Y_a \times Y_b}{Y_w} \quad (6)$$

where $X$ is export volume, $Y$ is the gross domestic product and $a$, $b$, and w indices are respectively indicators of exporting country, importing country, and the world. As obvious in Equation (6), the coefficient of the fraction is equal to one, i.e. if this is the true model, estimated $\alpha$ in regression in Equation (5) will be not be significantly different from one.

Evenett and Keller (2002) give gravity equations in form of Equation 2 and Equation 3 based on two different imperfect specializations:

$$X_{ab} = (1-\gamma_a)\frac{Y_a \times Y_b}{Y_w} \quad (3)$$

$$X_{ab} = (\gamma_b - \gamma_a)\frac{Y_a \times Y_b}{Y_w} \quad (4)$$

In which $\gamma$ is the specialization index (a number between 0 and 1). As obvious, the coefficient of the fraction in these models is less than one. Evenett and Keller have shown that this coefficient is indeed less than one in bilateral trade data. They conclude thus that perfect specialization models are incapable of explaining the data.

The second criterion in Evenett and Keller (2002) is that the model should provide reasons why the coefficient departs from 1. They run different regressions[1] to estimate the coefficient of the fraction from data using five different levels of specialization, and claim that specialization index they use correlates reversely with the estimates of coefficient of the fraction. The results are summarized in panels (a) to (d) of Figure 1, and show a weak relationship.

## 3. A Model of Gravity with Perfect Specialization based on Tradable/Non-tradable Product Distinction

Assume that there are three commodities in world named as $s$, $t$ and $z$. Assume that s is not tradable and so to be precise we should use different notation for product s of each country. Assuming that there are two countries a and b, we call the s produced and consumed in country a: $s_a$, and the $s$ produced and consumed in country b: $s_b$. Either reason of perfect specialization, namely IRS forces or H-O model forces, can be used in model. Perfect specialization leads to each country to produce either of t or z. We assume that $a$ is producing $t$ and $b$ is producing $z$.

So these countries GDP's are:

$$Y_a = t + s_a \quad (7)$$

$$Y_a = t + s_a \quad (8)$$

---

[1] All regressions use the same data, and Evenett and Keller do not provide any information on how these coefficients differ.

Assuming $\lambda$'s denotes tradable share of GDP we have:

$$\lambda_a = \frac{t}{Y_a} = \frac{t}{t+s_a} \qquad (9)$$

$$\lambda_b = \frac{z}{Y_b} = \frac{z}{z+s_b} \qquad (10)$$

Supposing identical homothetic preferences, we have each country's share in consumption of each commodity is equal to its share of world GDP, i.e.

$$M_a = \frac{Y_a}{Y_w} z = \lambda_b \frac{Y_a \times Y_b}{Y_w} \qquad (11)$$

To simplify the perfect specialization model, Evenett and Keller (2002) assume that the share of non-tradable goods in GDP is identical for all countries. We do not simplify further as we believe that the idea of tradable share of GDP plays a great role in forming gravity equation and any simplification might lead to unreasonable results. Thus our model leads in a gravity equation with a coefficient of the ratio less than 1. To support the second criteria we shall show that the deviation of estimate of coefficient of the ratio is related to the share of non-tradable production in GDP. To show this we take logarithms of Equation (11) to get:

$$\ln(X_{ab}) = \beta_0 + \beta_1 \ln(\lambda_a) + \beta_2 \ln(Y_a \times Y_b / Y_w) + \varepsilon \qquad (12)$$

which is in fact the logarithm of Equation (11):

$$X_{ab} = e^{\beta_0} + \lambda_a^{\beta_1} + (Y_a \times Y_b / Y_w)^{\beta_2} \qquad (13)$$

Thus $\hat{\beta}_1 = \hat{\beta}_2 = 1$ shows accordance of model to data. On the other hand tradable production share in GDP is important only if $\hat{\beta}_1$ has a significant coefficient.

## 4. Data
World commodity trade data are gathered from UN ComTrade and UN Service Trade dataset provides the data on trade of services. National account data are from World Development Indicators dataset. 40 countries are selected that constitute a large part of world GDP (about 90%) and world trade. Saudi Arabia and Israel are dropped because of technical problems such as missing data. Table 2 lists the countries used in this study. The data form a panel of 1600 (40x40) export relationship over 10 years (2000-2009).

### 4.1. Calculating Tradable Productions Share
Model proposed in this paper is based on the tradable production share in GDP, so we shall provide some data on this phenomenon. But in reality no data are gathered for tradability. We only observe traded goods and services not what was potentially tradable. We study sectors of production and compare the shares of each sector in world production and world trade and decide if that section is tradable. For example agriculture constitutes about 5.61 percent of the world trade, but only 3.35 percent of world GDP, thus we can say that agriculture is tradable. Adding up tradable sectors we can calculate the tradability index of each country. Yet this calculation is not precise because of absolute decision on tradability of sectors. Textiles' share, as an example, is less than 0.5 percent of world GDP, yet more than 6 percent in world trade, so in fact textile is much more tradable than agriculture. So we use relative tradability with comparing the shares with the most tradable sector of the economy (textiles). So if 100 percent of textile is considered tradable, 81% of chemical and 12.36% of agricultural products are tradable. On the other hand only 2.31 percent of services (which is considered a non-tradable sector) are tradable.

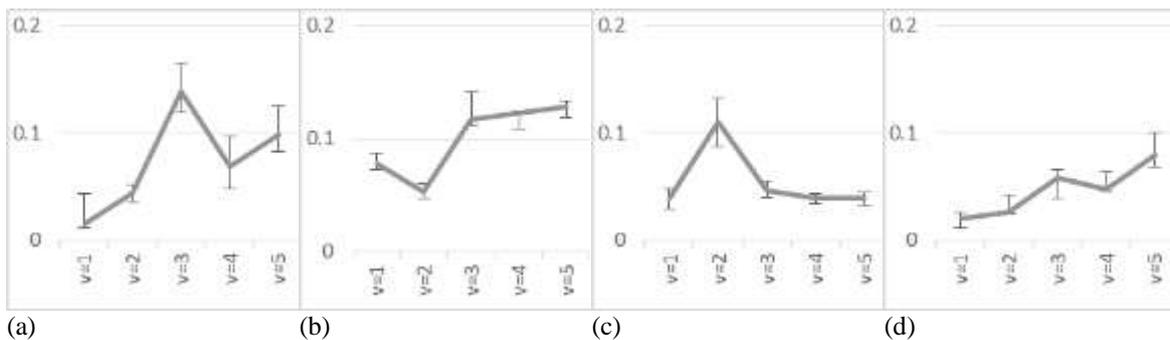

(a)      (b)      (c)      (d)

**Figure (1): Estimates of Coefficients of Gravity Equation versus Specialization Level**

*Source*: Evenett and Keller (2002)

**Table 1: Tradability of Each Economic Sector**

| Sector | Share in (%) | | Ratio | Result | Relative Tradability |
|---|---|---|---|---|---|
| | World GDP | World Trade | | | |
| Agriculture | 3.35 | 5.61 | 1.67 | Tradable | 12.36 |
| Non-manufacturing industry | 10.00 | 7.20 | 0.72 | Non-tradable | 5.32 |
| Chemicals | 1.68 | 18.40 | 10.95 | Tradable | 80.84 |
| Food, beverages and tobacco | 1.87 | 1.12 | 0.60 | Non-tradable | 4.42 |
| Machinery and transport equipment | 4.57 | 32.99 | 7.21 | Tradable | 53.25 |
| Other manufacturing | 7.77 | 7.08 | 0.91 | Non-tradable | 6.73 |
| Textiles and clothing | 0.46 | 6.17 | 13.55 | Tradable | 100.00 |
| Services | 68.37 | 21.43 | 0.31 | Non-tradable | 2.31 |

*Source*: Authors

We use the data from Table 1 to calculate the tradability index for each country for each year. Table 2 reports the average tradability index for countries of the study from 2000 to 2009. As shown in this table, Indonesia, Singapore, Malaysia, and South Korea have the highest indices. This means that these countries have production structures that are able to export more. The calculated index is not based on trade data of these countries and is only based on the production structure. Data in Table 2, is plotted on the map of the world in Figure 2.

## 5. The Results

Equation 10 is estimated on the bilateral trade data which is a 1560x10 panel. This panel is unbalanced by nature (not all countries have exported to all countries in every year). 6624 observations are available. We estimated the equation using both fixed effect model and random effect model. Testing the null hypothesis that no panel effect exits (thus recommending use of pooled estimates) is rejected. Hausman (1978) test indicates that panel effects are fixed effects, and random effects estimations leads to biased estimates of the equation (Baltagi 2008). Table 3 reports the results of the both models and Hausman test results are reported in Table 4. As we can see in Table 3, both coefficients of interest $\beta_1$ and $\beta_2$ are statistically not different than 1, thus the theoretic model is supported with data. $\hat{\beta}_2 = 1$ means that the core part of the gravity equation, i.e. that trade is positively related to the multiplication of GDP of both partners is modeled in a way that is completely compatible with data. $\hat{\beta}_1 = 1$ means that the tradability index is the sole reason for deviations of data from basic perfect specialization models.

## 6. Conclusion

We provided a perfect specialization model based on the tradability phenomenon, which does not have the problems indicated by Evenett and Keller (2002), namely that our perfect specialization model completely explains the deviations of data from simpler perfect specialization models without entrapment in the complexities of imperfect specialization models (which we do not believe are doing any good in explaining the data). Empirical evidence using estimations on panel data of bilateral trade of 40 countries over 10 years completely and fully supports the theoretical model. In the process of providing empirical evidence, we built and reported an index of tradability which is a measure of the potentials of a country to be an exporter. The results showed that tradability was the merely reason for deviations of data from basic perfect specialization models.

**References**

1. Bergstrand, J. H. (1989), "The Generalized Gravity Equation, Monopolistic Competition and the Factor-Proportions Theory in International Trade," *The Review of Economics and Statistics*, 71(1): 143-153.
2. Bohman H. and D. Nilson (2007), "Market Overlap and the Direction of Exports: A New Approach of Assessing the Linder Hypothesis," Working Paper Series in Economics and Institutions of Innovation. 86. Royal Institute of Technology.
3. Deardorff, A. (1998), Determinants of Bilateral Trade: Does Gravity Work in a Neoclassical World?, in The Regionalization of the World Economy, edited by J.A. Frankel. Chicago: Univ. Chicago Press for National Bureau of Economic Research. 7-32.
4. Evenett, S. J. and W. Keller (2002), "On Theories Explaining the Success of the Gravity Equation," *Journal of Political Economy*, 110(2), 281-361.

5. Feenstra, R.C., J. A. Markusen and A. K. Rose (1998), Understanding the Home Market Effect and the Gravity Equation: The Role of Differentiating Goods (No. w6804), National Bureau of Economic Research.
6. Hausman, J.A. (1978), "Specification Tests in Econometrics, *Econometrica,* 46(6), 1251-1271.
7. Helpman, E. and P.R. Krugman (1985), Market Structure and Foreign Trade: Increasing Returns, Imperfect Competition, and the International Economy, MIT Press.
8. Krugman, P.R., and M. Obstfeld (2009), International Economics: Theory and Policy, Boston Pearson Addison-Wesley, The Pearson Series in Economics .
9. Linder, S.B. (1961), an Essay on Trade and Transformations, Stockholm: Almqvist & Wiksell, New York, J. Wiley.
   Ricardo, D. (1817), the Principles of Political Economy and Taxation, London: John Murray, Albemarle-Street.

Table 2: Average of 2000-2009 Tradability Index for 40 Countries

| Country | Tradability | Country | Tradability |
|---|---|---|---|
| Argentina | 7.31 | Japan | 9.57 |
| Australia | 4.83 | Malaysia | 13.56 |
| Austria | 7.77 | Mexico | 8.12 |
| Belgium | 8.03 | Netherlands | 6.27 |
| Brazil | 8.25 | Norway | 5.93 |
| Canada | 7.56 | Poland | 6.86 |
| China | 6.19 | Portugal | 7.45 |
| Colombia | 8.00 | Rep. of Korea | 13.24 |
| Czech Rep. | 8.92 | Russian Federation | 6.70 |
| Denmark | 5.27 | Singapore | 13.67 |
| Egypt | 11.24 | South Africa | 7.02 |
| Finland | 9.21 | Spain | 7.50 |
| France | 7.04 | Sweden | 7.63 |
| Germany | 9.99 | Switzerland | 6.99 |
| Greece | 5.25 | Thailand | 10.59 |
| India | 10.35 | Turkey | 10.60 |
| Indonesia | 13.74 | United Arab Emirates | 11.94 |
| Iran | 8.45 | United Kingdom | 6.69 |
| Ireland | 11.40 | USA | 7.10 |
| Italy | 9.13 | Venezuela | 11.76 |

*Source*: Authors

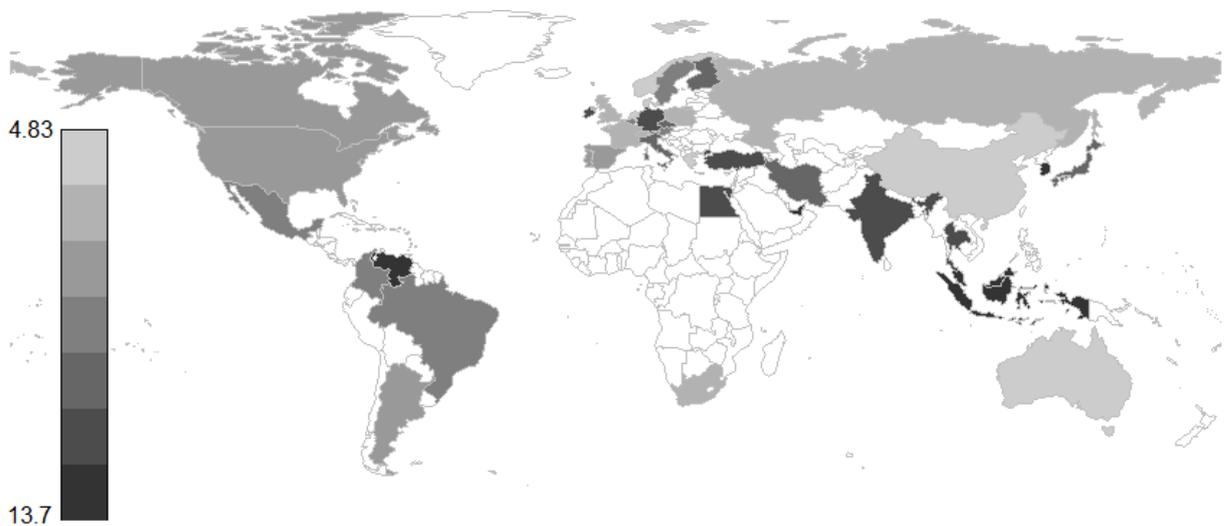

**Figure (2): Average of 2000-2009 Tradability Index for 40 Countries**
*Source*: Authors

**Table 3: Estimating the Gravity Equation based on Imperfect Specialization Model of Bilateral Trade; Fixed and Random Effect Models**

| Coefficient | Fixed Effects Model | Random Effects Model |
|---|---|---|
| $\hat{\beta}_0$ | -4.4434*** | -3.7589*** |
|  | (0.3798) | (0.3354) |
| $\hat{\beta}_1$ | 0.9573*** | 0.7117*** |
|  | (0.1039) | (0.0861) |
| $\hat{\beta}_2$ | 1.0178*** | 1.0098*** |
|  | (0.0130) | (0.3355) |
| No. of Observations | 6624 | 6624 |
| F Test (Degrees of Freedom) | 3077.41 |  |
| Degrees of Freedom of F Test | 2, 5067 |  |
| Prob > F | 0.0000 |  |
| Wald Chi-squared Test |  | 7523.48 |
| Degree of Freedom of Wald Test |  | 2 |
| Prob > Chi2 |  | 0.0000 |

*Source*: Authors